\documentclass[longauth]{aa} 
\usepackage{graphicx}
\usepackage{txfonts}
\usepackage[colorlinks=true,linkcolor=red,citecolor=blue,filecolor=black,urlcolor=blue]{hyperref}

\begin{document} 

    \title{Uncovering the stellar structure of the dusty star-forming galaxy GN20 at z=4.055 with MIRI/JWST}

   \author{L. Colina\inst{\ref{inst:CAB}} \and A. Crespo G\'omez\inst{\ref{inst:CAB}} \and J. \'Alvarez-M\'arquez\inst{\ref{inst:CAB}}  \and A. Bik\inst{\ref{inst:Stockholm}} \and F. Walter\inst{\ref{inst:MPIA}} \and L. Boogaard\inst{\ref{inst:MPIA}} \and A. Labiano\inst{\ref{inst:CAB},\ref{inst:Telespazio}} \and  F. Peissker\inst{\ref{inst:Kln}}
   \and P. P\'erez-Gonz\'alez\inst{\ref{inst:CAB}} \and G. {\"O}stlin\inst{\ref{inst:Stockholm}} \and T.R. Greve\inst{\ref{inst:DTU}, \ref{inst:DAWN}, \ref{inst:UCL}} \and H.U. N{\o}rgaard-Nielsen\inst{\ref{inst:DTU}} 
   \and G. Wright\inst{\ref{inst:UKATC}} 
   \and A. Alonso-Herrero\inst{\ref{inst:CAB-ESAC}} \and R. Azollini\inst{\ref{inst:CAB}, \ref{inst:Dublin}} \and K.I. Caputi\inst{\ref{inst:Groningen},\ref{inst:DAWN}} 
   \and D. Dicken\inst{\ref{inst:UKATC}}
   \and M. Garc\'ia-Mar\'in\inst{\ref{inst:ESA}} 
   \and J. Hjorth\inst{\ref{inst:DARK}} 
   \and O. Ilbert\inst{\ref{inst:LAM}} 
   \and S. Kendrew\inst{\ref{inst:ESA}}  
   \and J.P. Pye\inst{\ref{inst:Leicester}} \and T. Tikkanen\inst{\ref{inst:Leicester}} 
   \and P. van der Werf\inst{\ref{inst:Leiden}} 
   \and L. Costantin\inst{\ref{inst:CAB}} \and E. Iani\inst{\ref{inst:Groningen}} \and S. Gillman\inst{\ref{inst:DTU}, \ref{inst:DAWN}} \and I. Jermann\inst{\ref{inst:DTU}, \ref{inst:DAWN}} \and D. Langeroodi\inst{\ref{inst:DARK}}  \and T. Moutard\inst{\ref{inst:LAM}}  \and P. Rinaldi\inst{\ref{inst:Groningen}} \and M. Topinka\inst{\ref{inst:Dublin}}
   \and E.F. van Dishoeck\inst{\ref{inst:Leiden}}  
   \and M. G\"udel\inst{\ref{inst:Vienna}, \ref{inst:MPIA}, \ref{inst:ETH}} \and Th. Henning\inst{\ref{inst:MPIA}} \and P.O. Lagage\inst{\ref{inst:AIM}} \and T. Ray\inst{\ref{inst:Dublin}} \and B. Vandenbussche\inst{\ref{inst:Leuven}}} 

   \institute{Centro de Astrobiolog\'{\i}a (CAB), CSIC-INTA, Ctra. de Ajalvir km 4, Torrej\'on de Ardoz, E-28850, Madrid, Spain\\  \email{jalvarez@cab.inta-csic.es} \label{inst:CAB}
   \and Department of Astronomy, Stockholm University, Oscar Klein Centre, AlbaNova University Centre, 106 91 Stockholm, Sweden \label{inst:Stockholm}
    \and Max-Planck-Institut f\"ur Astronomie, K\"onigstuhl 17, 69117 Heidelberg, Germany\label{inst:MPIA}
    \and Telespazio UK for the European Space Agency, ESAC, Camino Bajo del Castillo s/n, 28692 Villanueva de la Ca\~{n}ada, Spain \label{inst:Telespazio}
    \and I.Physikalisches Institut der Universit\"at zu K\"oln, Z\"ulpicher Str. 77, 50937 K\"oln, Germany \label{inst:Kln}
    \and DTU Space, Technical University of Denmark, Elektrovej 327, 2800 Kgs. Lyngby, Denmark \label{inst:DTU}
    \and Cosmic Dawn Centre (DAWN), Copenhagen, Denmark \label{inst:DAWN}
    \and Department of Physics and Astronomy, University College London, Gower Place, London WC1E 6BT, UK \label{inst:UCL}
    \and UK Astronomy Technology Centre, Royal Observatory Edinburgh, Blackford Hill, Edinburgh EH9 3HJ, UK \label{inst:UKATC} 
    \and Centro de Astrobiolog\'ia (CAB), CSIC-INTA, Camino Viejo del Castillo s/n, 28692 Villanueva de la Ca\~{n}ada, Madrid, Spain \label{inst:CAB-ESAC}  
    \and Dublin Institute for Advanced Studies, Astronomy \& Astrophysics Section, 31 Fitzwilliam Place, Dublin 2, Ireland \label{inst:Dublin}
    \and Kapteyn Astronomical Institute, University of Groningen, P.O. Box 800, 9700 AV Groningen, The Netherlands \label{inst:Groningen}   
    \and European Space Agency, Space Telescope Science Institute, Baltimore, Maryland, USA \label{inst:ESA} 
    \and DARK, Niels Bohr Institute, University of Copenhagen, Jagtvej 128, 2200 Copenhagen, Denmark \label{inst:DARK} 
     \and Aix Marseille Universit\'e, CNRS, LAM (Laboratoire d’Astrophysique de Marseille) UMR 7326, 13388, Marseille, France \label{inst:LAM} 
    \and School of Physics \& Astronomy, Space Research Centre, Space Park Leicester, University of Leicester, 92 Corporation Road, Leicester, LE4 5SP, UK \label{inst:Leicester} 
    \and Leiden Observatory, Leiden University, PO Box 9513, 2300 RA Leiden, The Netherlands \label{inst:Leiden}              
    \and Centre for Extragalactic Astronomy, Durham University, South Road, Durham DH1 3LE, UK \label{inst:Durham}
    \and University of Vienna, Department of Astrophysics, Türkenschanzstrasse 17, 1180 Vienna, Austria \label{inst:Vienna}
    \and Institute of Particle Physics and Astrophysics, ETH Zurich, Wolfgang-Pauli-Str 27, 8093 Zurich, Switzerland \label{inst:ETH} 
    \and AIM, CEA, CNRS, Universit\`e Paris-Saclay, Universit\`e Paris Diderot, Sorbonne Paris Cit\`e, F-91191 Gif-sur-Yvette, France \label{inst:AIM}
    \and Institute of Astronomy, KU Leuven, Celestijnenlaan 200D bus 2401, 3001 Leuven, Belgium \label{inst:Leuven}   
    }

   \date{Received ; accepted}


 

\abstract 
{Luminous infrared galaxies at high redshifts ($z$\,>\,4) include extreme starbursts that build their stellar mass over short periods of time, that is, of 100 Myr or less. These galaxies are considered to be the progenitors of massive quiescent galaxies at intermediate redshifts ($z$\,$\sim$\,2) but their stellar structure and buildup is unknown. Here, we present the first spatially resolved near-infrared (rest-frame 1.1\,$\mu$m) imaging of GN20, one of the most luminous dusty star-forming galaxies known to date, observed at an epoch when the Universe was only 1.5\,Gyr old. The 5.6\,$\mu$m image taken with the JWST Mid-Infrared Instrument (MIRI/JWST) shows that GN20 is a very luminous galaxy (M$_\mathrm{1.1\,\mu m,\,AB}$\,=\,$-$25.01, uncorrected for internal extinction), with a stellar structure composed of a conspicuous central source and an extended envelope. The central source is an unresolved nucleus that carries 9\% of the total flux. 
The nucleus is co-aligned with the peak of the cold dust emission, and offset by 3.9\,kpc from the ultraviolet stellar emission. The diffuse stellar envelope is similar in size (3.6\,kpc effective radius) to the clumpy CO molecular gas distribution. The centroid of the stellar envelope is offset by 1\,kpc from the unresolved nucleus, suggesting GN20 is involved in an interaction or merger event supported by its location as the brightest galaxy in a proto-cluster. Additional faint stellar clumps appear to be associated with some of the UV- and CO-clumps.
The stellar size of GN20 is larger by a factor of about 3 to 5 than known spheroids, disks, and irregulars at $z$ $\sim$ 4, while its size and low S\'ersic index are similar to those measured in dusty, infrared luminous galaxies at redshift 2 of the same mass ($\sim$\,10$^{11}$\,M$_{\odot}$). GN20 has all the ingredients necessary for evolving into a massive spheroidal quiescent galaxy at intermediate redshift: it is a large, luminous galaxy at $z$\,=\,4.05 involved in a short and massive starburst centred in the stellar nucleus and extended over the entire galaxy, out to radii of 4 kpc,  and likely induced by the interaction or merger with a member of the proto-cluster.}

\titlerunning{MIRI/JWST study of GN20}
\maketitle

\section{Introduction}
\label{sec:1.}

High-redshift ($z$\,>\,2\,$-$\,3) infrared-bright (IR-bright) galaxies represent the dusty star formation phase of the stellar buildup and galaxy assembly \citep{Casey+14,Lutz+14}. Recent deep ALMA surveys have concluded that IR-bright sources dominated star formation in the Universe up to $z$\,$\sim$\,4, contributing 35\% at $z$\,$\sim$\,5, and even  20\,$-$\,30\% at $z$\,$\sim$\,6\,$-$\,7 \citep{Gruppioni+20,Zavala+21,Algera+23}, representing an important fraction of the star formation (SF) in the early Universe that has been missing from the deepest rest-frame optical and ultraviolet (UV) surveys \citep{Bouwens+22}. A fraction of these IR-luminous galaxies, extreme starbursts, or dusty star-forming galaxies (DSFGs, hereafter), show short (<\,100 Myr), intense starburst episodes, forming stars at rates of  500\,$-$\,1000\,M$_{\odot}$\,yr$^{-1}$, and higher, namely, at the expected maximal starburst rate \citep{Thompson+05, Crocker+18, Walter+22}. These DSFGs are the most luminous starbursts in the Universe and are considered the progenitors of massive quiescent galaxies at redshifts $z$\,$\sim$\,2\,$-$\,3 \citep{Toft+14}. The study of the stellar structure in these early DSFGs is fundamental for consolidating our understanding of the formation and stellar buildup of massive galaxies. However, due to their faintness at (rest-frame) optical wavelengths, the stellar light distribution of the host galaxies is unknown. Yet these data are now achievable with the JWST's exquisite combination of sensitivity and sub-arcsec angular resolution \citep{Rigby+23}.

This letter presents an analysis of new mid-infrared imaging of GN20 obtained with the JWST Mid-Infrared Instrument (MIRI) \citep{Rieke+15, Wright+15}. GN20 is a DSFG at a redshift of 4.0554 \citep{Carilli+11} located in a proto-cluster or galaxy overdensity  \citep{Daddi+09}. GN20, identified as a bright 850\,$\mu$m source in GOODS$-$North \citep{Pope+06}, has an infrared luminosity ($L_\mathrm{IR}$) of 1.86\,$\times$\,10$^{13}$\,L$_{\odot}$, and a star formation rate (SFR) of 1860\,M$_{\odot}$\,yr$^{-1}$, assuming a Chabrier IMF \citep{Tan+14}. The molecular gas distribution shows a clumpy structure with a diameter of 14\,kpc \citep{Hodge+12} and a kinematics that is consistent with that of a massive ($M_\mathrm{dyn}$\,=\,5.4\,$\pm$\,2.4\,$\times$\,10$^{11}$\,M$_{\odot}$) rotating disk. Finally, PAH emission at 6.2\,$\mu$m was detected in GN20, representing the first detection of PAHs at a redshift above $z$\,=\,4 \citep{Riechers+14}. 

Section~\ref{sec:2.} introduces the JWST and ancillary data, describing the (post-)calibration processing of the JWST imaging. Section~\ref{sec:3.} discusses the stellar structure of the galaxy ($\S$\ref{sec:3.1.}), a comparison with the rest-frame UV and far-infrared continuum and the cold molecular gas ($\S$\ref{sec:3.2.}), a consideration of its starburst nature  ($\S$\ref{sec:3.3.}) and how it compares to the general galaxy population at a redshift of 4 ($\S$\ref{sec:3.4.}).  A summary of our findings is given in $\S$\ref{sec:4.}. Throughout this paper, we assume a Chabrier initial mass function \citep{Chabrier+03}, a flat $\Lambda$CDM cosmology with $\Omega_\mathrm{m}$\,=\,0.310, and H$_0$\,=\,67.7\,km\,s$^{-1}$\,Mpc$^{-1}$ \citep{PlanckCollaboration18VI}. For this cosmology, 1 arcsec corresponds to 7.08\,kpc at $z$\,=\,4.05 and the luminosity distance is $D_\mathrm{L}$\,=\,37.13\,Gpc.

\section{Observations, calibration, and data processing}
\label{sec:2.}

\subsection{JWST MIRI data and calibration}
\label{sec:2.1.}

GN20 JWST imaging was obtained on November 23-24, 2022 using the  MIRI imager \citep[MIRIM,][]{Bouchet+15} in the F560W filter as part of the European Consortium MIRI Guaranteed Time (program ID 1264). The observation has a total integration time of 1498.5 seconds using the FASTR1 read-out mode and a five-dither medium-size cycling pattern, with one integration of 108 groups per dither. The MIRIM F560W image has been calibrated using version 1.9.5 of the JWST pipeline and context 1077 of the Calibration Reference Data System (CRDS). The  process follows the same steps applied in the calibration of the SPT0311-58 image (\citealt{Alvarez-Marquez+23}). A final image with a scale of 0.06$\arcsec$ per pixel and 0.24\arcsec{} FWHM is used throughout the analysis. At the redshift of GN20, the MIRI F560W image traces the rest-frame 1.1$\mu$m emission.

\subsection{Ancillary imaging}
\label{sec:2.2.}

The HST WFC3/F105W calibrated image of GN20 was retrieved from the Mikulski Archive for Space Telescopes (PI: Faber, ID: 12442). This image was taken with an integration time of $\sim$\,2800\,s, drizzled to a pixel scale of $\sim$\,0.09\arcsec{} and with a resolution FWHM\,$\sim$\,0.24\arcsec. Archival Plateau de Bure Interferometer (PdBI) tuned to 340\,GHz (i.e. 880\,$\mu$m), and Very Large Array (VLA) imaging of the CO(2-1) line with a final resolution of $\sim$\,0.19\arcsec{} \citep{Hodge+15} is also included in the analysis. Details about these observations and their data processing can be found in \citet{Carilli+11} and \citet{Hodge+12}.

\subsection{JWST-VLA-HST astrometry}
\label{sec:2.3.}

An absolute positioning better than 100\,mas is required for ensuring a proper comparison of the GN20 structure traced with the multi-wavelength high angular resolution imaging. The WFPC3/HST and MIRI/JWST absolute astrometry was derived by measuring the centroid of the positions of stars in the field of view with available GAIA DR3 \citep{GaiaCollaboration+22} coordinates, giving an uncertainty of less than $\sim$\,70\,mas in their astrometry. The VLA and PdBI observations were phase-referenced to quasars, and the absolute astrometry is accurate to a tenth of the synthesised beam, namely, about 20\,mas. The astrometric uncertainties that result when comparing MIRI structures with those identified in the VLA, PdBI, and HST data are therefore less than 70\,mas. 

\begin{figure*}
\centering
   \includegraphics[width=0.85\linewidth]{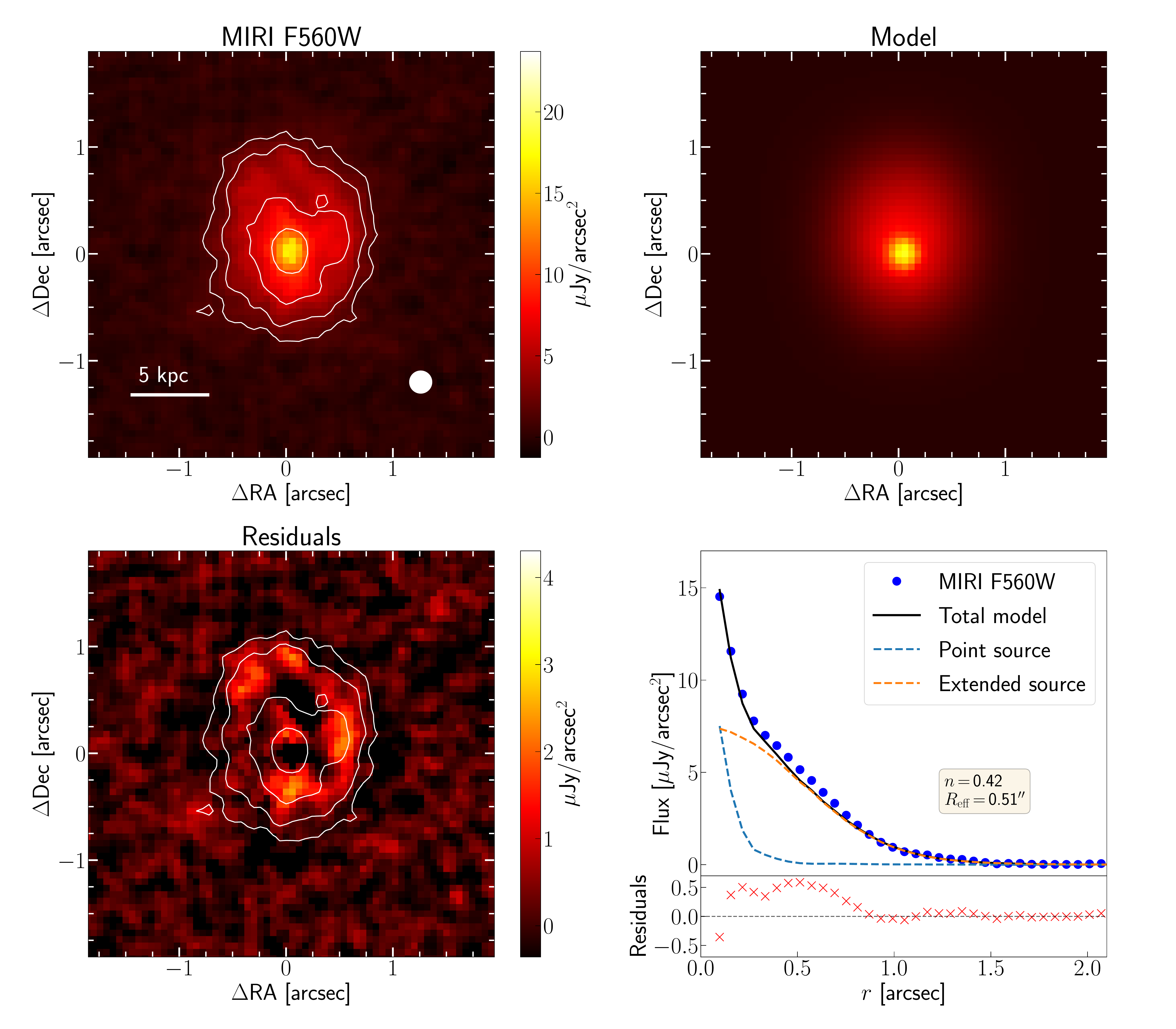}
      \caption{Two-dimensional, two-component (PSF + S\'ersic) fit to the GN20 light distribution. The panels represent the original MIRI F560W image (top left), the best-fit model obtained with \texttt{Lenstronomy} (top right), and its residuals (bottom left), respectively. Bottom right panel shows the radial profile of the original image (blue dots), best-fit model (black line), and residuals (red crosses) maps. In addition, blue and orange dashed lines represent the radial profiles of the nuclear point-source and extended components used in the modelling of the light distribution. Contours represent the F560W isophotes at 5$\sigma$, 10$\sigma$, 20$\sigma,$ and 35$\sigma$ levels.}
         \label{fig:lenstronomy_fit}
\end{figure*}

\section{Results and discussion}
\label{sec:3.}

\subsection{The stellar structure of GN20}
\label{sec:3.1.} 

The rest-frame 1.1$\mu$m stellar structure of GN20 shows a well resolved, extended emission of about 2\arcsec{} in diameter and with a bright central source (see Fig.~\ref{fig:lenstronomy_fit}). The 5.6$\mu$m flux measured at r\,$\sim$\,1.2\arcsec{}, which defines the 3$\sigma$ level, is 13.3\,$\pm$\,0.1\,$\mu$Jy, in good agreement with previous IRAC 5.8$\mu$m flux (14.3\,$\pm$\,0.7\,$\mu$Jy; \citealt{Barro+19}). To quantify the properties of the stellar light distribution, we performed parametric fits on the F560W light distribution using the \texttt{Lenstronomy} code \citep{Birrer+18}, allowing us to perform an MCMC analysis to estimate the associated uncertainties. An empirical PSF created from the two Gaia stars in the MIRI FoV was used during this modelling. For comparisons with previous studies, we first consider a single component fit using a S\'ersic model. This analysis yields a S\'ersic index of $n$\,=\,0.64\,$\pm\,$0.02,  $R_\mathrm{eff}$\,=\,3.40\,$\pm$\,0.02\,kpc, and an axial ratio of $b/a$\,=\,0.81\,$\pm$\,0.01. The high residuals observed in the central region when applying this single S\'ersic fit motivates a two-component decomposition (see Fig.~\ref{fig:lenstronomy_fit}). This approach indicates the presence of a central point source (stellar nucleus, hereafter) and an extended disk (stellar envelope hereafter), characterised by a surface brightness distribution with $R_\mathrm{eff}$\,=\,3.60\,$\pm$\,0.03\,kpc, $n$\,=\,0.42\,$\pm$\,0.02, and $b/a$\,=\,0.80\,$\pm$\,0.01. Furthermore, the centroid of this extended emission is offset by 0.14 arcsec (i.e. 1\,kpc) from the nuclear point source. For this second approach, we obtain smaller $\chi^2$ and BIC values, validating the use of the two-component fit (see \citealt{Liddle+07}). The same analysis has been performed using \texttt{GALFIT} \citep{Peng+02,Peng+10} to test the self-consistency of the results, obtaining values that are similar. The modelled fluxes of the point and extended sources represent the 9\% (1.15\,$\pm$\,0.05\,$\mu$Jy) and 85\,\% (11.25\,$\pm$\,0.11\,$\mu$Jy) of the total flux, respectively, while the other 6\,\% is linked to the stellar clumps visible in the residuals (Fig.~\ref{fig:lenstronomy_fit}).

The centroid of the stellar envelope is offset from the stellar nucleus by a distance of 1\,kpc. Non-concentric isophotes are predicted as a result of tidal interactions \citep{Aguilar+86} and have been detected in the brightest galaxies in clusters at low-redshift \citep{Lauer+88}. This is a transient, short-lived distortion due to the dynamical friction of the stellar nucleus with the envelope. GN20 is the brightest member in a galaxy overdensity with two other DSFGs and one LBG spectroscopically confirmed at a redshift of 4.05, and an  additional 11 B-dropout galaxies, all at  distances of less than 25\arcsec{} \citep{Daddi+09}. Its stellar structure could therefore be affected by tidal forces due to the interaction with other close galaxies. An alternative is that GN20 is in a late-merger state as the potential presence of a secondary nucleus is identified (see inset in Fig.~\ref{fig:gn20_images}, top-right panel) when applying a Lucy$-$Richardson \citep{Lucy+74} deconvolution of the F560W image with the empirical PSF plus a two-pixel Gaussian Kernel filter \citep{Peissker+22}. 

\subsection{Star formation in GN20. The UV, near-IR, molecular gas and cold dust perspectives}
\label{sec:3.2.} 

\begin{figure*}
\centering
   \includegraphics[width=0.85\linewidth]{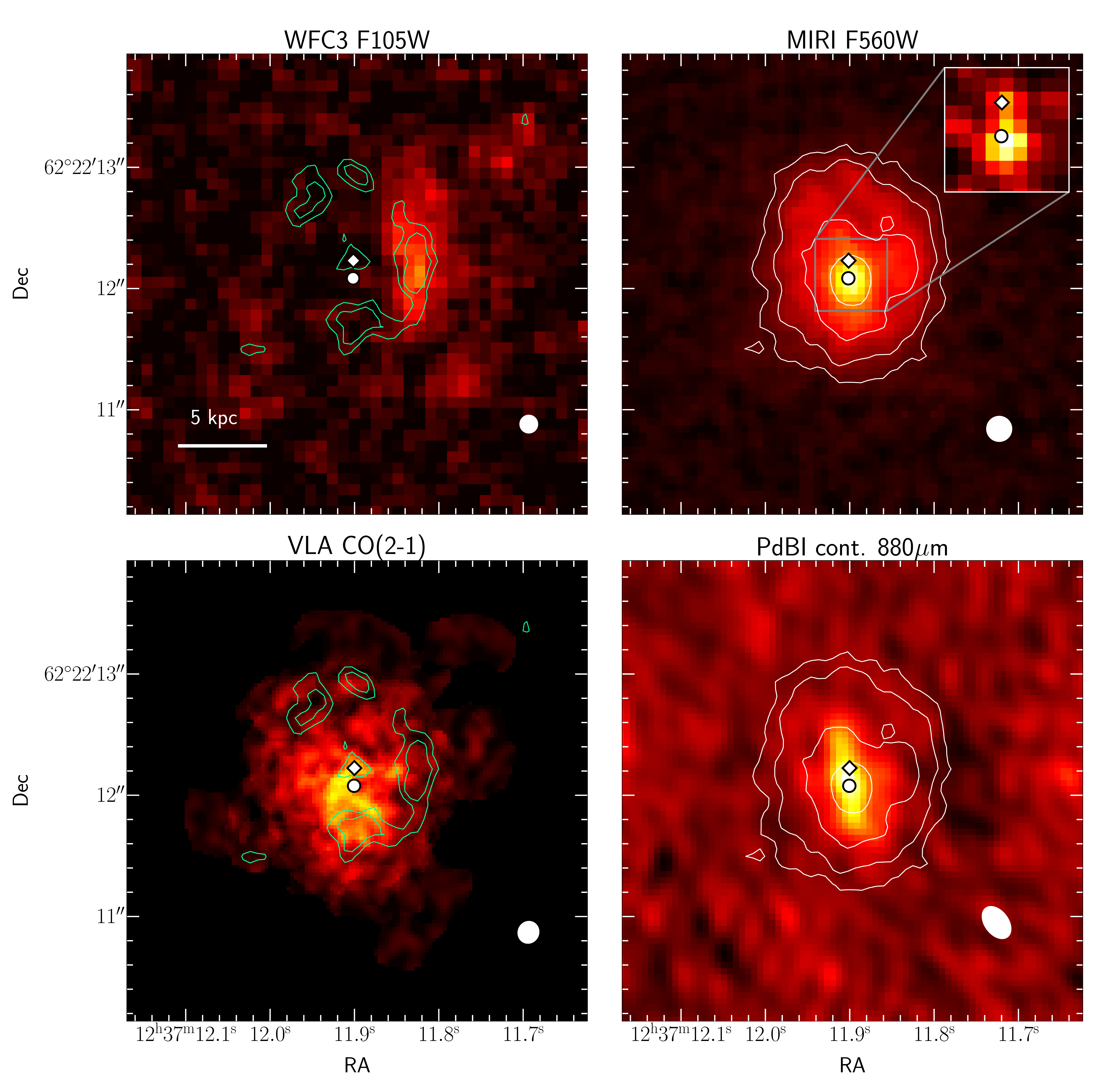}
      \caption{Multi-wavelength morphology of GN20. Top panels display the WFC3 F105W (left) and MIRI F560W (right) images of GN20, tracing the rest-frame UV (0.2$\mu$m) and near-IR (1.1$\mu$m) light, respectively. Bottom left panel shows the CO(2-1) flux map, obtained with the VLA. Bottom right panel displays the rest-frame 170\,$\mu$m continuum map from PdBI observations. Black-edged white diamond and circle mark the position of the centre for the nuclear point-source and extended components derived from the 2D brightness decomposition (see Sect.~\ref{sec:3.1.}). White circles and ellipses at the bottom right corner of the panels represent the PSF or beam size for the different images. White contours on the right panels represent the F560W isophotes at 5$\sigma$, 10$\sigma$, 20$\sigma,$ and 35$\sigma$ levels. Green contours in the left panels mark the residuals of the near-IR light distribution fit (see Fig.~\ref{fig:lenstronomy_fit}) at 3$\sigma$ and 5$\sigma$ levels. The inset in the F560W image represents the central 0.6\arcsec\,$\times$\,0.6\arcsec{} region with the presence of the secondary nucleus as obtained from the deconvolution of the F560W image (see Sect.~\ref{sec:3.1.})}
         \label{fig:gn20_images}
\end{figure*}

\begin{figure}
\centering
   \includegraphics[width=\linewidth]{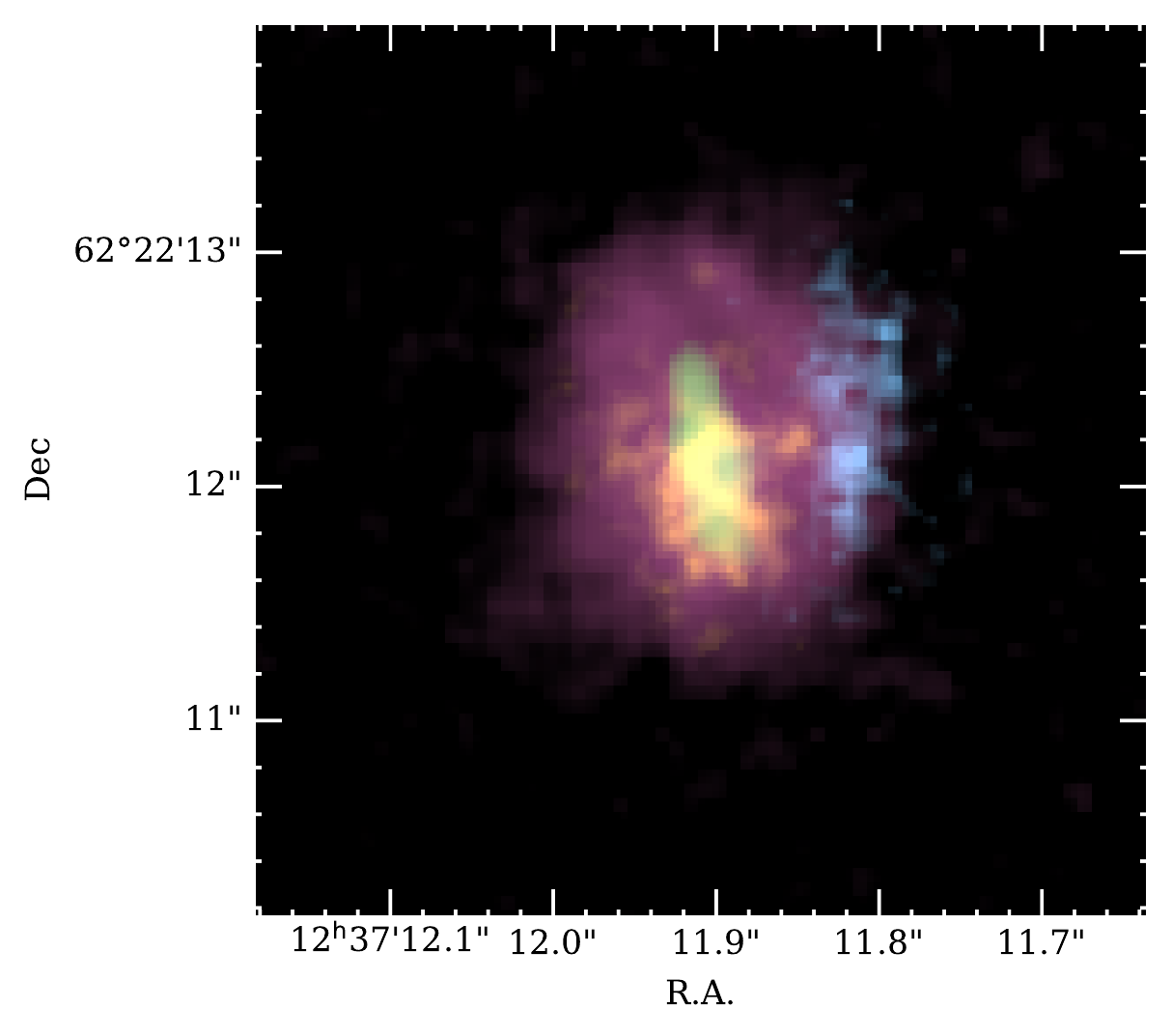}
      \caption{Composite image of GN20 displaying the UV emission (HST/F105W, blue), cold dust continuum (PdBI/880$\mu$m, lime), molecular gas (VLA CO(2-1), orange), and stellar (MIRI/F560W, purple) components of the galaxy.}
         \label{fig:rgb}
\end{figure}

GN20 has been previously imaged with HST and radio interferometers at a similar angular resolution as the MIRI imaging. HST (rest-frame) 0.2$\mu$m imaging traces the unobscured, young star formation. VLA and PdBI imaging \citep{Hodge+12, Hodge+15} trace the cold molecular gas and dust emission, respectively (see Figs.~\ref{fig:gn20_images} and~\ref{fig:rgb}). The stellar structure traced by the F560W/MIRI image combined with the ancillary data provide a unique picture of this DSFG with several new results:
1) there is a substantial offset of 0.55\arcsec{} (i.e. 3.9\,kpc) between the stellar nucleus and the UV-emitting regions. These UV-bright regions are located W-NW of the nucleus, in an arc-like structure in the outskirts of the galaxy; 2) the stellar nucleus perfectly coincides with the position of the far-infrared continuum emission peak;  3) the stellar nucleus is offset by 0.14\arcsec{} (i.e. 1\,kpc) west of the CO(2-1) emission peak; 4) the centroid of the stellar envelope is offset by 0.14\arcsec{} north of the far-infrared continuum and by 0.20\arcsec{} (1.4\,kpc) northwest of the CO(2-1) peak emission; 5) the clumpy molecular gas is embedded in the stellar envelope and is similar in size; and 6) the stellar clumps (traced by the residuals in the two-component light decomposition, (see $\S$\ref{sec:3.1.} and Fig.~\ref{fig:lenstronomy_fit}) coincide with either the UV-emitting or some of the molecular clumps (Fig.~\ref{fig:gn20_images}). 

The picture that emerges from the combination of the multi-wavelength imaging is quite enlightening. The stellar nucleus appears to have the highest concentration of cold dust but not the largest concentration of molecular gas. It appears with a lower molecular gas content relative to the circumnuclear molecular clumps. There is also a clear confirmation that the known large offset between the UV-bright emitting regions and the cold dust and molecular gas distribution must be due to a large dust obscuration within the central few kpc in the galaxy as previously identified in other high$-z$ dusty star-forming galaxies \citep{Hodge+16, Gomez-Guijarro+18}. Under the hypothesis that the CO(2-1) and the UV emitting regions trace the obscured and unobscured star formation, respectively, GN20  already appears as a large galaxy (see $\S$\ref{sec:3.4.}) where  star formation is proceeding at all scales from the nucleus (<\,1\,kpc) to the circumnuclear regions (1\,$-$\,3\,kpc), and (to a lesser degree) in the more external regions of the stellar envelope (>\,3\,kpc). 

\citealt{Cortzen+20} have argued that the dust emission in GN20 is optically thick up to (rest-frame) 170\,$\mu$m. Their best model requires the dust to be concentrated in the nuclear region with $R_\mathrm{eff}$\,=\,1.2\,kpc, reaching mass surface densities ($\Sigma_\mathrm{dust}$) of 500\,M$_{\odot}$\,pc$^{-2}$. The upper size of this extreme, dust-enshrouded region is smaller than the size of the extended stellar envelope by a factor of 3
 and, therefore, it does not  affect the overall 1.1\,$\mu$m light distribution. However, it is known that low$-z$ (ultra)luminous infrared galaxies do show a clumpy dust distribution at (sub)kpc scales, with the highest concentration in the nucleus and decreasing outwards (e.g. \citealt{Piqueras-Lopez+13, Gimenez-Arteaga+22}). A similar dust distribution in GN20 will affect the surface brightness of the stellar structure with more prominent clumps and patchy dust-lanes in the ultraviolet and optical, while showing a more diffuse emission and bright nucleus in the near-infrared. Future approved JWST imaging with NIRCam will cover the rest-frame optical wavelengths and, in combination with the HST and MIRI data, will provide the detailed two-dimensional (sub)kpc structure of the dust distribution and extinction effects in GN20.

\subsection{Nature of the nuclear source and extended emission in GN20} 
\label{sec:3.3.} 

The unresolved stellar nucleus (upper limit on the size of 0.8\,kpc) is very luminous with an (uncorrected by internal extinction) absolute magnitude (M$_\mathrm{1.1\,\mu m,\,AB}$) of $-$22.35. This high luminosity could be due to the presence of a massive bulge, an AGN, or an obscured nuclear starburst. Under the hypothesis of a bulge, the rest-frame 1.1\,$\mu$m luminosity traces stellar mass and would translate to a mass of 2.5\,$\times$\,10$^{10}$\,M$_{\odot}$ for an intermediate-age (300 Myr) old stellar population, according to STARBURST99 \citep{Leitherer+99}. Extinction-corrected luminosities and older populations would imply even larger masses as the $M/L$ ratio increases with the age of the population. Compact massive galaxies at redshifts of 4 and with masses of up to 10$^{11}$\,M$_{\odot}$ are known to exist \citep{Valentino+20}. However, these galaxies are quiescent, that is, their SFR is well below that of the average main-sequence of star-forming galaxies \citep{Speagle+14}, while GN20 is in a starburst phase (as derived from its infrared luminosity). If the observed near-infrared stellar structure of GN20 is due to an interaction or advanced merger (see $\S$\ref{sec:3.1.}), dusty nuclear starbursts accompanied by AGNs should be present \citep{Ricci+17, Blecha+18}. Some evidence for a heavily obscured AGN comes from the mid- and far-infrared. The ratio of the detected PAH6.2$\mu$m to the infrared luminosity and the upper limit of the X-ray luminosity to the 6$\mu$m luminosity is compatible with the presence of an obscured, Compton-thick, but bolometrically-weak AGN, carrying only 1\% of GN20 bolometric luminosity \citep{Riechers+14}. Therefore, the most likely nature of the nucleus is a dust-enshrouded starburst as also supported by the recent claim that the dust emission is optically thick up to 170\,$\mu$m \citep{Cortzen+20}. The mean depletion time in GN20 derived from the 880\,$\mu$m continuum and CO(2-1) emission is 130 Myr with average star formation surface densities of 100\, M$_{\odot}$\,yr$^{-1}$\,kpc$^{-2}$ and nuclear peak of 119 M$_{\odot}$\,yr$^{-1}$\,kpc$^{-2}$ \citep{Hodge+15}. For a constant star formation rate over a period of 100 Myr, the SFR in the unresolved stellar nucleus derived from its (extinction uncorrected) near-infrared luminosity, is 45\,M$_{\odot}$\,yr$^{-1}$, namely, at the lower limit to previous values. The size and SFR in the nucleus of GN20 is also similar to that of compact starbursts identified in $z$ $\sim$3-6 sub-millimetre galaxies (SMGs) with ALMA \citep{Ikarashi+15}. However, GN20 shows in addition evidence of forming stars also in the extended stellar envelope ($\S$\ref{sec:3.2.}). Assuming the 1.1$\mu$m luminosity in the stellar envelope is due to a starburst with properties equal to that of the nucleus, a total (nucleus and envelope) SFR and stellar mass of 524\,M$_{\odot}$\,yr$^{-1}$ and 5.2\,$\times$\,10$^{10}$\,M$_{\odot}$ were derived. These (extinction-uncorrected) values represent lower limits to the SFR and stellar mass derived from the overall SED and IR luminosity (1860\,M$_{\odot}$\,yr$^{-1}$ and 1.1\,$\times$\,10$^{11}$\,M$_{\odot}$, \citealt{Tan+14}). Thus, the new MIRI imaging confirms the classification of GN20 as an extended dusty starburst with a specific star formation (sSFR) of about 10\,$-$\,17\,Gyr$^{-1}$, well above the main-sequence of star-forming galaxies at $z$\,=\,4 \citep{Speagle+14}.

\subsection{GN20 and the general population of high-z galaxies} 
\label{sec:3.4.}

\begin{figure*}
\centering
   \includegraphics[width=0.85\linewidth]{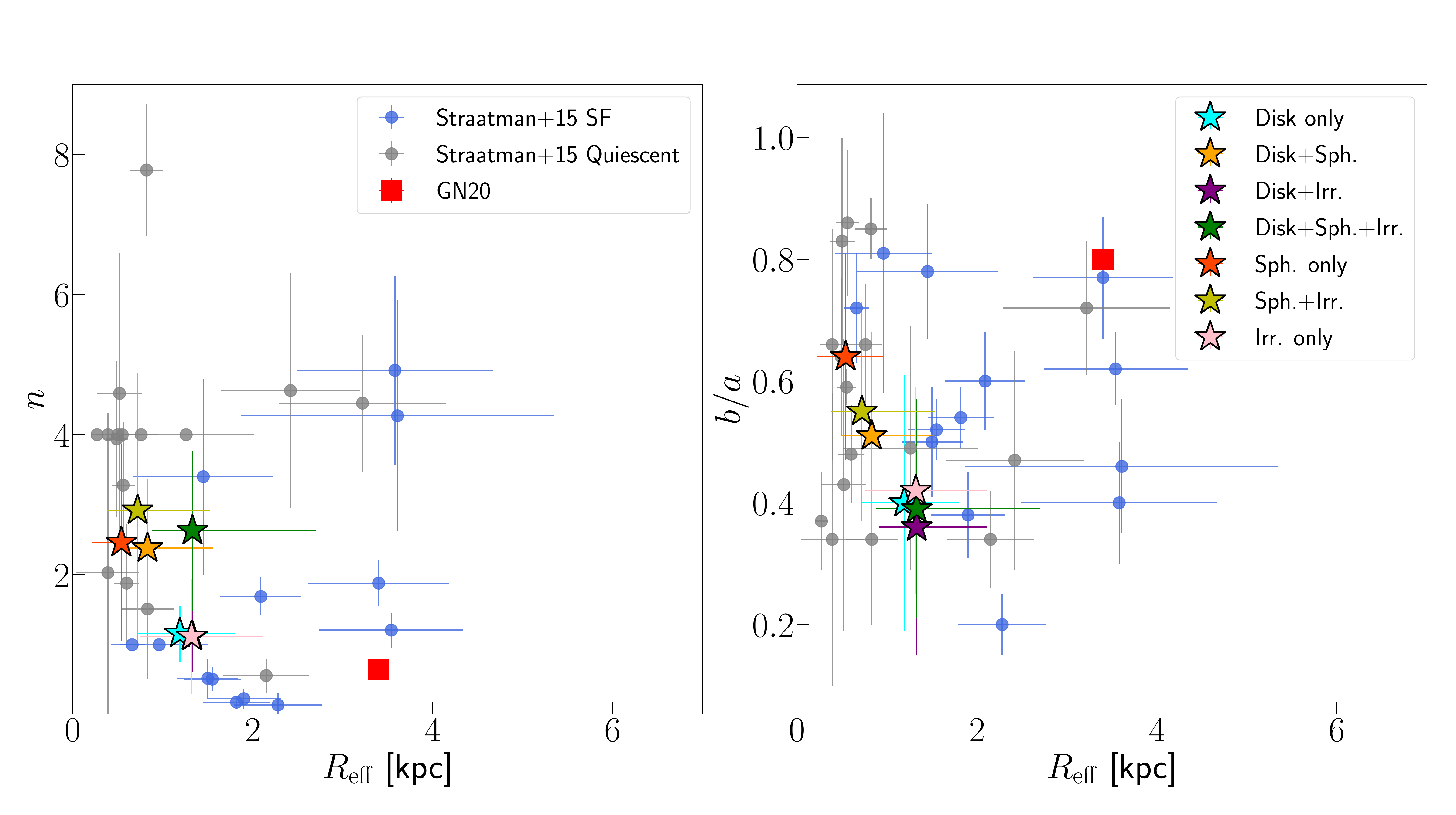}
      \caption{Structural parameters, $R_\mathrm{eff}-n$ (left) and $R_\mathrm{eff}-b/a$ (right), for different populations of $z$\,$\sim$\,4 galaxies. Blue and grey circles represent the massive star-forming and quiescent galaxies, respectively, from the ZFOURGE and CANDELS surveys (Table\,1 of \citealt{Straatman+15}). Coloured stars display the mean values for each one of the morphologically-classified groups of CEERS galaxies (Table\,1 of \citealt{Kartaltepe+23}). The red squares are the values for GN20 and their sizes include their associated errors.}
         \label{fig:comparison}
\end{figure*}

Recent studies of the structure and morphology of the general population of galaxies at $z$\,>\,3 have   used CEERS NIRCam/JWST imaging \citep{Kartaltepe+23}. This study traces, for the first time, the rest-frame red light ($\sim$\,0.7\,$\mu$m) in high-z galaxies, covering all morphological types from spheroids to disks and irregulars. GN20 stands out with an effective radius that is larger than that of disks and spheroids at the average redshifts of 3.84 and 3.94 by a factor of
3 and 5, respectively. The S\'ersic index ($n$) has a low value of 0.64, within the range measured in galaxies classified as disks and irregulars (median $n$\,=\,1.16$^{+0.88}_{-0.40}$ and \,1.12$^{+1.51}_{-0.82}$, respectively, see Figure~\ref{fig:comparison}). Finally, the round shape of GN20 ($b/a$\,=\,0.81) has been attributed to its low inclination of 30\,$\pm$\,15 degrees \citep{Hodge+12}. The stellar light distribution in GN20 thus presents  peculiar properties that are not compatible with any of the existing morphological types. While its S\'ersic index appears closer to that of irregulars, the size is larger than galaxies of any of the morphological types at $z$\,$\sim$\,4 (see Figure~\ref{fig:comparison}). However, GN20 properties are similar to those of z\,$\sim$\,2 IR-luminous galaxies. Recent NIRCam/JWST imaging of a few massive ($\log{M_{\star}}$/M$_{\odot}$\,=\,10.9\,$-$\,11.7) IR-luminous galaxies ($\log{L_\mathrm{IR}}$/L$_{\odot}$\,=\,11.9\,$-$\,12.6) at (photometric) $z$\,$\sim$\,1.5\,$-$\,2.4 show that some of these galaxies are also large ($R_\mathrm{eff}$ of several to 10\,kpc) and exhibit a low S\'ersic index  ($n$\,=\,0.3 to 0.8) \citep{Chen+22}.

The origin of the large size of GN20 relative to the general population at $z$\,$\sim$\,4 could therefore be due to the mass difference. While the average stellar mass of the CEERS galaxies is in the 1.6\,$-$\,3.1\,$\times$\,10$^9$\,M$_{\odot}$ range, the SED-based stellar mass for GN20 is 1.1\,$\times$\,10$^{11}$\,M$_{\odot}$ \citep{Tan+14}. Following the size-mass relations in high-z galaxies \citep{vanderWel+14}, effective radii of 3.7\,kpc and 1.6\,kpc are predicted for disks and spheroids that are of the same mass as GN20. The effective radius of GN20 is in good agreement with the value for disks, despite the difference in redshift. However, recent NIRCam synthetic imaging based on TNG50 cosmological simulations \citep{Costantin+23} predicted significantly smaller sizes (based on an effective radius of 1.6\,kpc) for GN20-mass galaxies at redshift 4. 
The morphological characteristics of GN20 may also be compared with those of massive (i.e. $\sim$\,0.4\,$-$\,1\,$\times$\,10$^{11}$\,M$_{\odot}$) quiescent and star-forming galaxies at (photometric) redshifts around 4 \citep{Straatman+15}. With the caveat that the stellar structure in these galaxies is traced at rest-frame 0.3\,$\mu$m, GN20 appears as an extreme case of a star-forming galaxy, even when compared with the known population of $z$\,$\sim$\,4 extended star-forming galaxies (see Figure~\ref{fig:comparison}). In summary, GN20 appears as extreme in its stellar structure, when compared with other z\,$\sim$\,4 galaxy populations, and closer to z\,$\sim$\,2 IR-luminous galaxies, as if GN20 had experienced an accelerated evolution. In this respect,
GN20 is the brightest galaxy in a region that exhibits a galaxy overdensity and is identified as a proto-cluster (see $\S$\ref{sec:3.1.}). It could well be that GN20 had experienced an early growth in mass and size due to previous mergers with other, less massive members of the proto-cluster .

\section{Summary}
\label{sec:4.}
This letter presents the first mid-infrared sub-arcsec 
imaging at 5.6\,$\mu$m of the dusty star-forming galaxy GN20 at a redshift of 4.05, taken with MIRI/JWST. The image resolves for the first time the stellar structure by tracing the rest-frame 1.1$\mu$m light on scales of 1.5\,kpc. The new MIRI imaging is combined with existing multi-wavelength ancillary data tracing the rest-frame UV continuum and the cold molecular gas and dust with similar angular resolutions. This dataset yields a new picture of the stellar structure, its relation with the active star-forming regions (both obscured and unobscured) and its potential evolution.  

GN20 is a luminous galaxy (M$_\mathrm{1.1\mu m,\,AB}$\,=\,$-$25.01 uncorrected by internal extinction) with a stellar structure characterised by a luminous unresolved ($<$\,0.8\,kpc) nucleus and a diffuse extended envelope. The nucleus carries 9\% of the total flux and coincides with the compact, cold dust nuclear emission, and it is 3.9\,kpc away from the ultraviolet light that traces unobscured recent star-formation. The stellar envelope is characterised by an effective radius of 3.6\,kpc, a low S\'ersic index (0.42), and axis ratio ($b/a$) of 0.8. The position and extent of the stellar envelope agrees with that of the CO(2-1) molecular gas, while its centroid is offset by 1\,kpc from the stellar nucleus. Since GN20 is located in a proto-cluster, this offset is interpreted as the result of a recent gravitational encounter or merger. Additional faint stellar clumps are associated with some of the UV- and CO-clumps.

GN20 is a large galaxy with already a well developed stellar structure forming new stars at a constant, high rate (extinction uncorrected SFR\,$\sim$\,500\,M$_\mathrm{\odot}$\,yr$^{-1}$ for a period of 100\,Myr) not only in its nucleus, but also within the main body of the galaxy -- and even in the most external regions (distances 4\,kpc from the nucleus). GN20 is a galaxy that is three to five  times larger than (less massive) disks, irregulars, and spheroids at $z$\,$\sim$\,4, and similar to some massive DSFGs at $z$\,$\sim$\,2. Assuming gas depletion times of about 100\,Myr, GN20 is confirmed as an extended starburst with an sSFR of no less than 10\,Gyr$^{-1}$, well above the main sequence of star-forming galaxies at $z$\,=\,4. The early growth in the mass and size in GN20 may be related to its position as the brightest galaxy in a galaxy overdensity (or proto-cluster) environment where interactions and mergers are favoured. GN20 has all the ingredients necessary to evolve into a massive quiescent galaxy at intermediate redshift: it is a galaxy with a massive starburst centrally concentrated, but spatially extended over several kpc, with a short depletion time of $\sim$\,100\,Myr. This massive starburst was likely triggered by interactions and/or mergers with other members of the known $z$\,=\,4.05 proto-cluster. Further JWST multi-wavelength deep imaging and spectroscopy of GN20 and neighbouring galaxies is required to provide constraints on the spatially resolved physical properties, such as internal extinction, ages of the stellar populations, and additional morphological features of interactions or mergers involving GN20 and neighbouring galaxies.

\begin{acknowledgements}

This letter is dedicated to the memory of our colleague Hans Ulrik N{\o}rgaard-Nielsen, MIRI European Danish coPI and co-lead of the European Consortium MIRI High-z team, R.I.P.
Thanks to Jackie Hodge for providing detailed information about the positional accuracy of the VLA observations. J.A-M., A.C-G., L.C., A.L. acknowledge support by grant PIB2021-127718NB-100, and P.G.P-G. by grant PGC2018-093499-B-I00 funded by MCIN/AEI/10.13039/501100011033. A.A-H. is supported by grant PID2021-124665NB-I00 from the Spanish Ministry of Science and Innovation/State Agency of Research MCIN/AEI/10.13039/501100011033 and by “ERDF A way of making Europe”. F.W. and B.L. acknowledge support from the ERC Advanced Grant 740246 (Cosmic\_Gas), A.B. \& G.Ö. acknowledge support from the Swedish National Space Administration (SNSA).O.I. acknowledges the funding of the French Agence Nationale de la Recherche for the project iMAGE (grant ANR-22-CE31-0007), J.H. and D.L. were supported by a VILLUM FONDEN Investigator grant (project number 16599). K.I.C. acknowledges funding from the Netherlands Research School for Astronomy (NOVA) and the Dutch Research Council (NWO) through the award of the Vici Grant VI.C.212.036. J.P.P. and T.T acknowledge financial support from the UK Science and Technology Facilities Council, and the UK Space Agency. T.P.R. would like to acknowledge support from the ERC under advanced grant 743029 (EASY). The Cosmic Dawn Center (DAWN) is funded by the Danish National Research Foundation under grant No. 140

The work presented is the effort of the entire MIRI team and the enthusiasm within the MIRI partnership is a significant factor in its success. MIRI draws on the scientific and technical expertise of the following organisations: Ames Research Center, USA; Airbus Defence and Space, UK; CEA-Irfu, Saclay, France; Centre Spatial de Liége, Belgium; Consejo Superior de Investigaciones Científicas, Spain; Carl Zeiss Optronics, Germany; Chalmers University of Technology, Sweden; Danish Space Research Institute, Denmark; Dublin Institute for Advanced Studies, Ireland; European Space Agency, Netherlands; ETCA, Belgium; ETH Zurich, Switzerland; Goddard Space Flight Center, USA; Institute d'Astrophysique Spatiale, France; Instituto Nacional de Técnica Aeroespacial, Spain; Institute for Astronomy, Edinburgh, UK; Jet Propulsion Laboratory, USA; Laboratoire d'Astrophysique de Marseille (LAM), France; Leiden University, Netherlands; Lockheed Advanced Technology Center (USA); NOVA Opt-IR group at Dwingeloo, Netherlands; Northrop Grumman, USA; Max-Planck Institut für Astronomie (MPIA), Heidelberg, Germany; Laboratoire d’Etudes Spatiales et d'Instrumentation en Astrophysique (LESIA), France; Paul Scherrer Institut, Switzerland; Raytheon Vision Systems, USA; RUAG Aerospace, Switzerland; Rutherford Appleton Laboratory (RAL Space), UK; Space Telescope Science Institute, USA; Toegepast- Natuurwetenschappelijk Onderzoek (TNO-TPD), Netherlands; UK Astronomy Technology Centre, UK; University College London, UK; University of Amsterdam, Netherlands; University of Arizona, USA; University of Cardiff, UK; University of Cologne, Germany; University of Ghent; University of Groningen, Netherlands; University of Leicester, UK; University of Leuven, Belgium; University of Stockholm, Sweden; Utah State University, USA. A portion of this work was carried out at the Jet Propulsion Laboratory, California Institute of Technology, under a contract with the National Aeronautics and Space Administration. We would like to thank the following National and International Funding Agencies for their support of the MIRI development: NASA; ESA; Belgian Science Policy Office; Centre Nationale D'Etudes Spatiales (CNES); Danish National Space Centre; Deutsches Zentrum fur Luft-und Raumfahrt (DLR); Enterprise Ireland; Ministerio De Econom\'ia y Competitividad; Netherlands Research School for Astronomy (NOVA); Netherlands Organisation for Scientific Research (NWO); Science and Technology Facilities Council; Swiss Space Office; Swedish National Space Board; UK Space Agency. 

This work is based on observations made with the NASA/ESA/CSA James Webb Space Telescope. Some data were obtained from the Mikulski Archive for Space Telescopes at the Space Telescope Science Institute, which is operated by the Association of Universities for Research in Astronomy, Inc., under NASA contract NAS 5-03127 for \textit{JWST}; and from the \href{https://jwst.esac.esa.int/archive/}{European \textit{JWST} archive (e\textit{JWST})} operated by the ESDC.


\end{acknowledgements}

\bibliographystyle{aa} 
\bibliography{bibliography} 

\end{document}